\documentstyle[11pt,newpasp,twoside,epsf]{article}
\begin{document}
\title{Interstellar Dust Scattering Properties}
 \author{Karl D.\ Gordon}
\affil{Steward Observatory, University of Arizona, Tucson, AZ 85721}

\begin{abstract} 
Studies of dust scattering properties in astrophysical objects with
Milky Way interstellar dust are reviewed.  Such objects are reflection
nebulae, dark clouds, and the Diffuse Galactic Light (DGL).  To ensure
their basic quality, studies had to satisfy four basic criteria to be
included in this review.  These four criteria significantly reduced
the scatter in dust properties measurements, especially in the case of
the DGL.  Determinations of dust scattering properties were found to
be internally consistent for each object type as well as consistent
between object types.  The 2175~\AA\ bump is seen as an absorption
feature.  Comparisons with dust grain models find general agreement
with significant disagreements at particular wavelengths (especially
in the far-ultraviolet).  Finally, unanswered questions and future
directions are enumerated.
\end{abstract}

\section{Introduction}

The scattering properties of dust provide a unique view into the
physical properties of dust grains.  This view is different from other
probes of dust grains (eg., extinction, polarization due to
differential extinction, abundances) as it allows their absorption and
scattering properties to be separated.  The scattering properties of
dust grains are indicators of the size and composition of dust grains.

The scattering of photons by dust grains can be described by a single
scattering albedo, $a$, and a scattering phase function,
$\Phi(\alpha)$, where $\alpha$ is the scattering angle.  In most
studies, $\Phi(\alpha)$ is approximated by the single parameter
Henyey-Green (1941) phase function
\begin{equation}
\Phi_{HG}(\alpha) = \frac{ \left(1 - g^2\right) }
   { 4\pi \left( 1 + g^2 + 2g\cos \alpha \right)^{3/2} }
\end{equation}
where
\begin{equation}
g = \int \Phi(\alpha) \cos\alpha~d\Omega = \left< \cos\alpha \right>.
\end{equation}
The $g$ parameter is referred to as the scattering phase function
asymmetry and varies from -1 (complete back scattering) to 0
(isotropic scattering) to 1 (complete forward scattering).  This
Henyey-Greenstein phase function is a good approximation for dust
grains, except possible in the far-ultraviolet (Witt 1989).  Other
analytic forms of the scattering phase function have been proposed
(e.g., Cornette \& Shanks 1992; Draine 2003b), but have yet to be used
in dust scattering studies.

There are three astrophysical objects which are suited for studies of
Milky Way interstellar dust scattering properties.  These are
reflection nebulae, dark clouds, and the Diffuse Galactic Light (DGL).
These objects consist of either high-density or diffuse dust
illuminated by sufficiently strong sources to permit the detection of
scattered light.  Reflection nebulae usually consist of a single or
small number of stars illuminating their natal cloud.  Dark clouds are
isolated dust clouds which are usually illuminated by the interstellar
radiation field.  The DGL is the sum of the light of the stars in the
Milky Way scattered by dust in the Galaxy.  Other objects which do
show scattering light, but which are not suitable for such studies are
circumstellar disks, \ion{H}{II} regions, and external galaxies.
These objects have dust which is newly formed, processed, or non-Milky
Way dust.

Each of these three astrophysical objects have different strengths and
weaknesses with respect to determining dust scattering properties.
The strengths of reflection nebulae are that they are bright, have
simple, wavelength independent illuminator geometries, and are
illuminated by a single or handful of stars.  Their weaknesses are
that they often have strong emissions in the red (Extended Red
Emission [ERE]) and near-infrared (non-equilibrium, thermal small
particle emission).  In addition, they usually probe high density
regions ($R_V > 3.1$) making the results harder to compare to most
dust grain models.  The strengths of dark clouds are that they have
simple, wavelength independent illuminator geometries and are
externally illuminated by the interstellar radiation field and/or a
small number of stars.  Dark clouds share the same weakness as
reflection nebulae as they also probe high density regions ($R_V >
3.1$) and, in addition, they are difficult to observe as they are
faint.  The strength of the DGL is that it probes the diffuse
interstellar medium ($R_V = 3.1$).  The weaknesses of the DGL are that
the illuminator geometry is wavelength dependent and the dust geometry
is complex having a disk density distribution with embedded dust
clouds.  In addition, DGL observations are challenging as observations
of the faint DGL signal are needed over a fairly large region of the
sky, especially a range of galactic latitudes, to allow for accurate
measurements of $a$ and $g$.  The combination of the results from all
three astrophysical objects should give a good view of dust scattering
properties without undue uncertainties resulting from each object
type's weaknesses.

This review will concentrate on published studies of interstellar dust
scattering properties in the ultraviolet, visible, and near-infrared.
The topic of X-ray scattering by dust is covered elsewhere in these
proceedings.  A great deal of work has taken place in this area since
the review of dust scattering properties at the last dust meeting
(Witt 1989).  At that time, only a handful of studies existed which
made quantitative determinations of dust scattering properties (ie.,
actually determined values for $a$ and $g$).  Since the last dust
meeting, a number of studies have specifically been addressed to this
issue.

Brief summaries of these studies, usually in the form of plots of $a$
and $g$ as a function of wavelength, have been presented in the
literature (eg., Gordon et al.\ 1994; Mathis 1996; Li \& Greenberg
1997; Witt \& Gordon 2000; Draine 2003a, 2003b).  One drawback to these
reviews has been the inclusion of all literature results without
consideration of the quality of each study.  This has lead to a
misplaced notion that our knowledge of dust scattering properties is
quite uncertain in some wavelength regions.  This has been especially
pronounced in studies of the DGL in the far-ultraviolet (around
1500~\AA).  In a handful of these studies, preliminary analyses have
been revised in later papers resulting in significantly higher
determinations of both $a$ and $g$.  Thus, it is important to apply a
uniform set of criteria when reviewing dust scattering studies to
ensure the basic quality of the results.  This has been done in
this review.

\section{Literature Results}

The literature was searched for determinations of dust scattering
properties in objects possessing Milky Way interstellar dust.
Astrophysical objects which satisfy this constraint are reflection
nebulae, dark clouds, and the DGL.

For inclusion in this review, each study had to satisfy four criteria.
These criteria are:
\begin{enumerate}

\item \emph{Determine albedo \& g}: This requires that specific values
be quoted for the two quantities.  Studies which were only sensitive
to one of the two quantities were included as long as a realistic
range of the other quantity was probed.  Usually, the albedo ($a$) was
the quantity determined with calculations performed for $g$ values
between zero and one.  While this criteria means that the large number
of early studies which only put limits on $a$ and/or $g$ are
neglected, their limits are usually consistent with the values
determined from later studies.

\item \emph{Determine uncertainties}: The importance of this criteria
cannot be understated.  The uncertainties presented here are either
the uncertainties quoted in the original paper or reflect the range of
$a$ \& $g$ values allowed by models presented in each study. This can
mean that multiple models (with different assumptions) have been
combined into a single measurement with larger uncertainties than a
single model.

\item \emph{Refereed Publication}:  This ensures that the full details
of the work have be presented. In addition to the usual journals (eg.,
\apj, \aj, \aap, \pasp, \mnras, etc.), theses were counted as refereed
publications.

\item \emph{Not be superseded by other work}: A number of studies are 
superseded by newer work which usually used better radiative transfer
models (especially the case for DGL studies).  This usually results in
significantly different determinations of $a$ \& $g$.

\end{enumerate}

\begin{figure}
\plotone{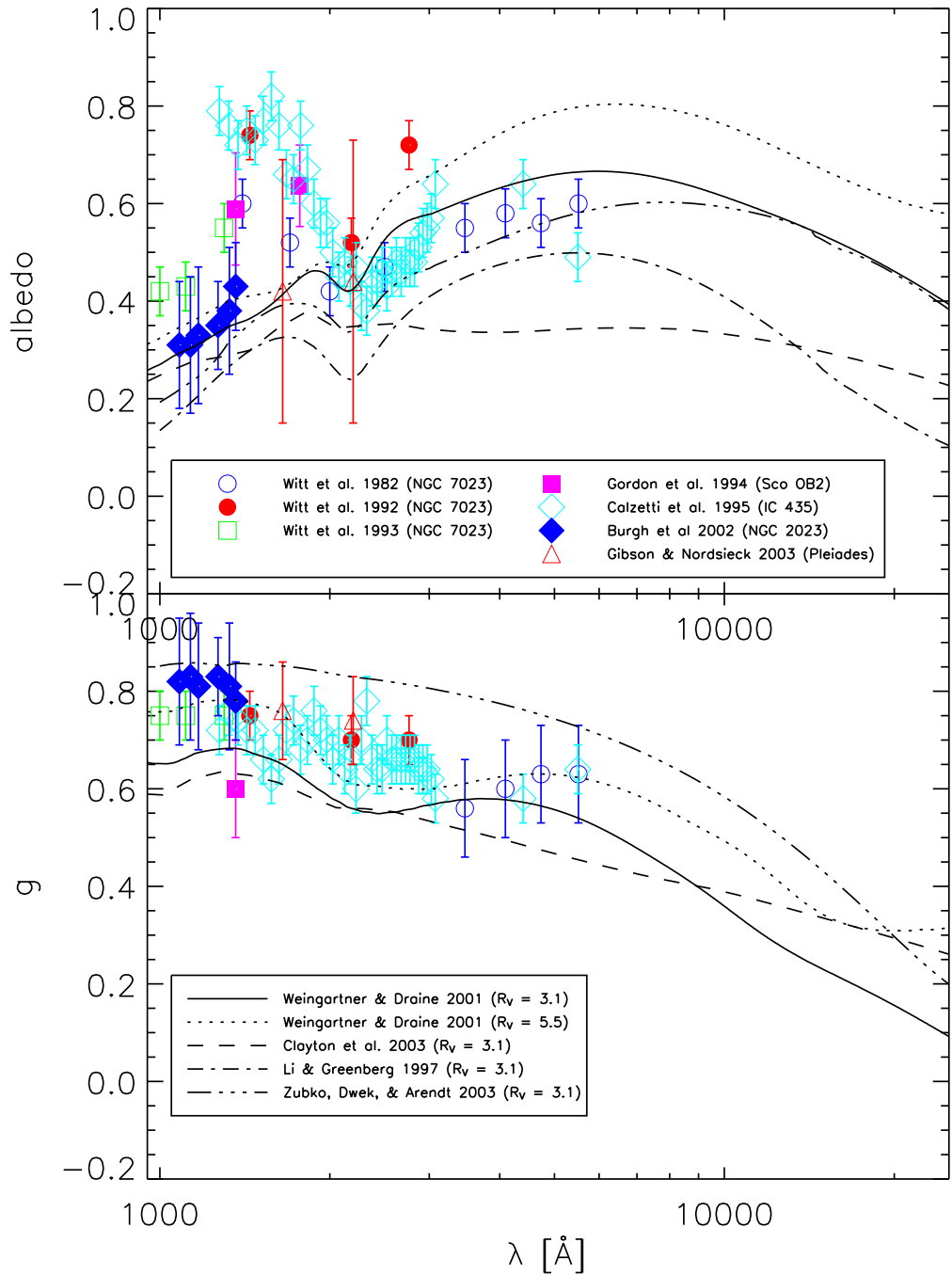}
\caption{Determinations of the albedo and g in reflection nebulae are
plotted versus wavelength.  In addition, predictions from dust grain
models are plotted for comparison. \label{fig_rn_vals}}
\end{figure}

\begin{figure}
\plotone{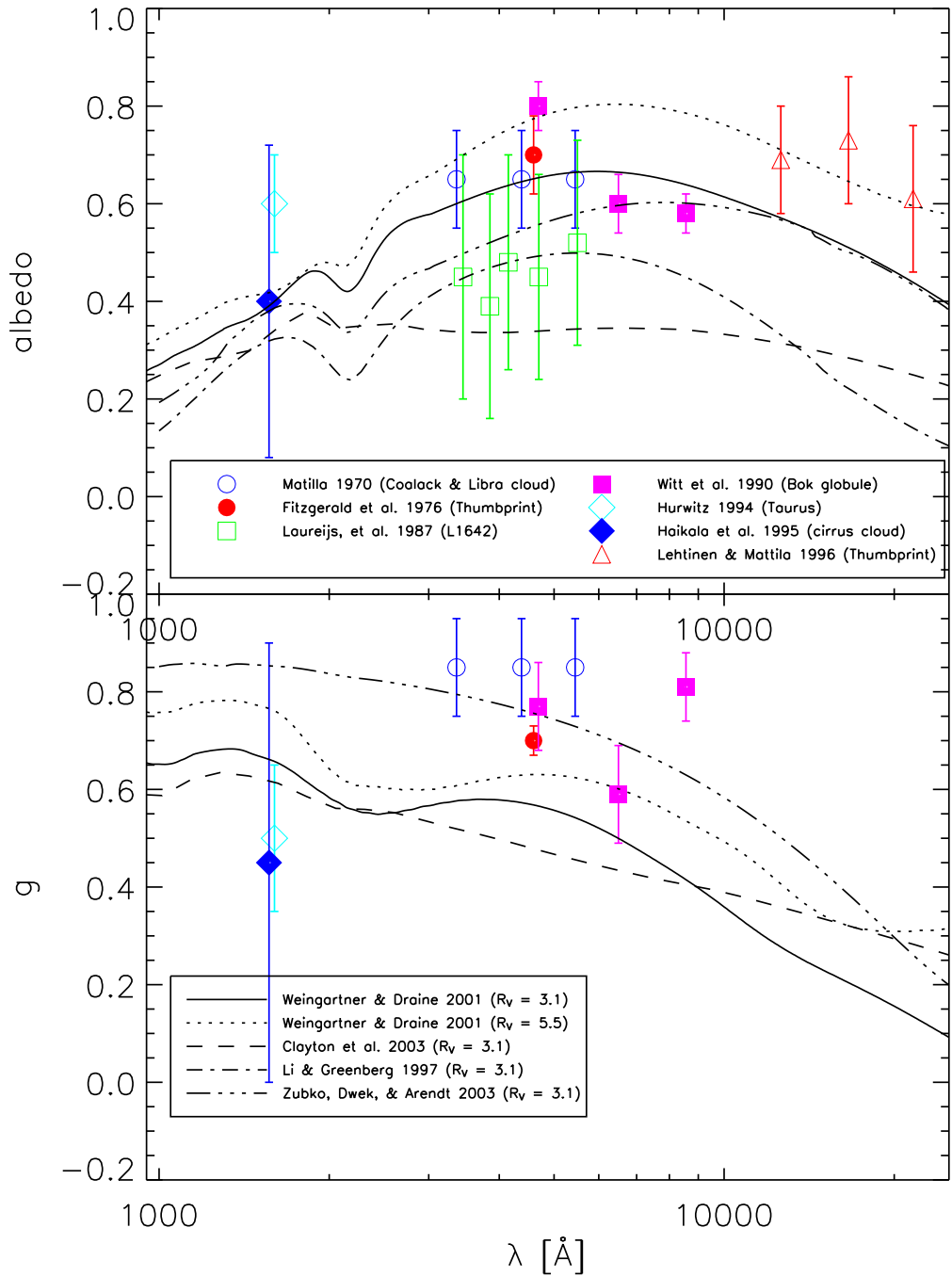}
\caption{Determinations of the albedo and g in dark clouds are plotted
versus wavelength.  In addition, predictions from dust grain models
are plotted for comparison. \label{fig_dc_vals}}
\end{figure}

\begin{figure}
\plotone{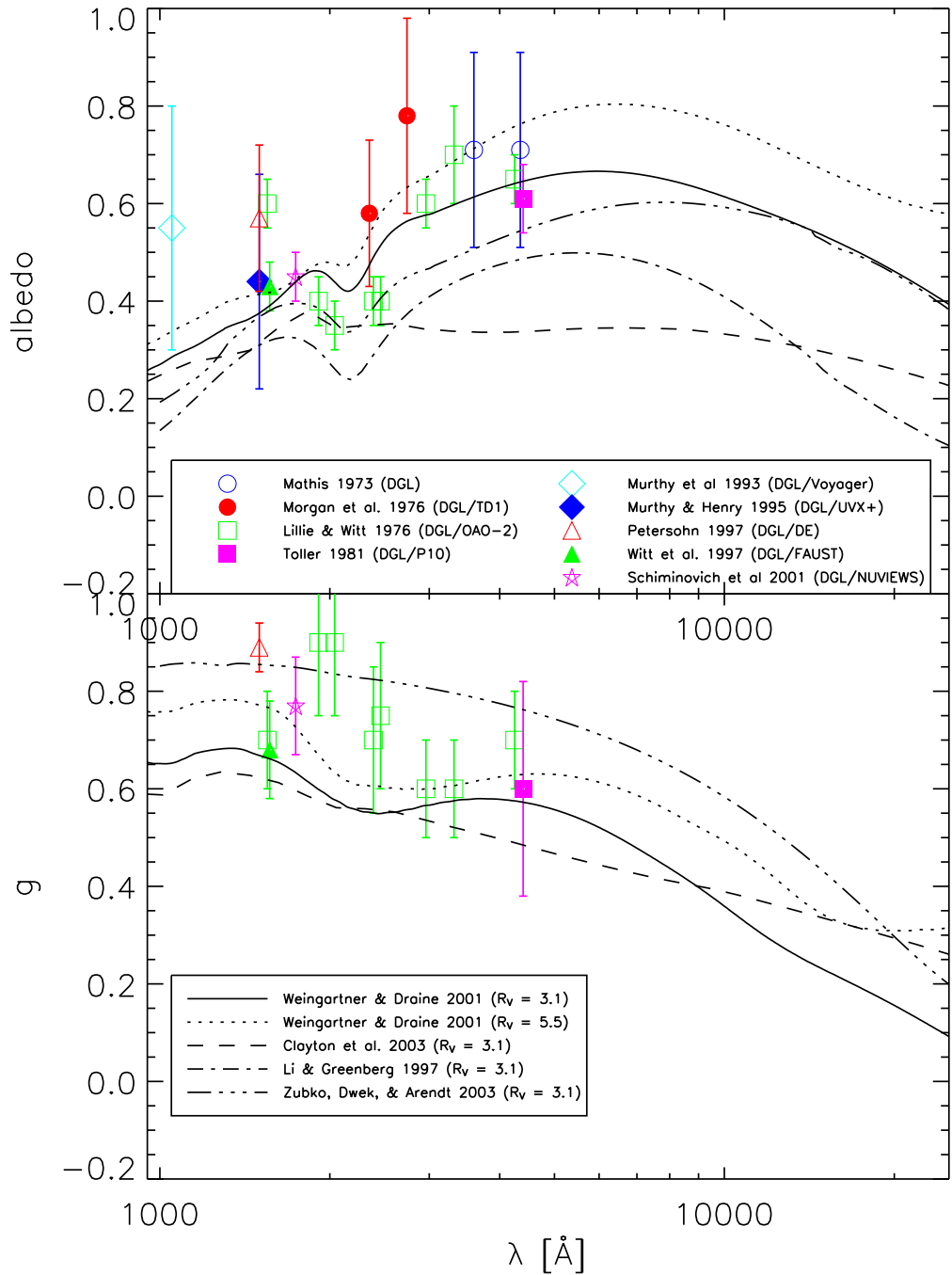}
\caption{Determinations of the albedo and g in the DGL are plotted
versus wavelength. In addition, predictions from dust grain models are
plotted for comparison.  \label{fig_dgl_vals}}
\end{figure}

\begin{figure}
\plotone{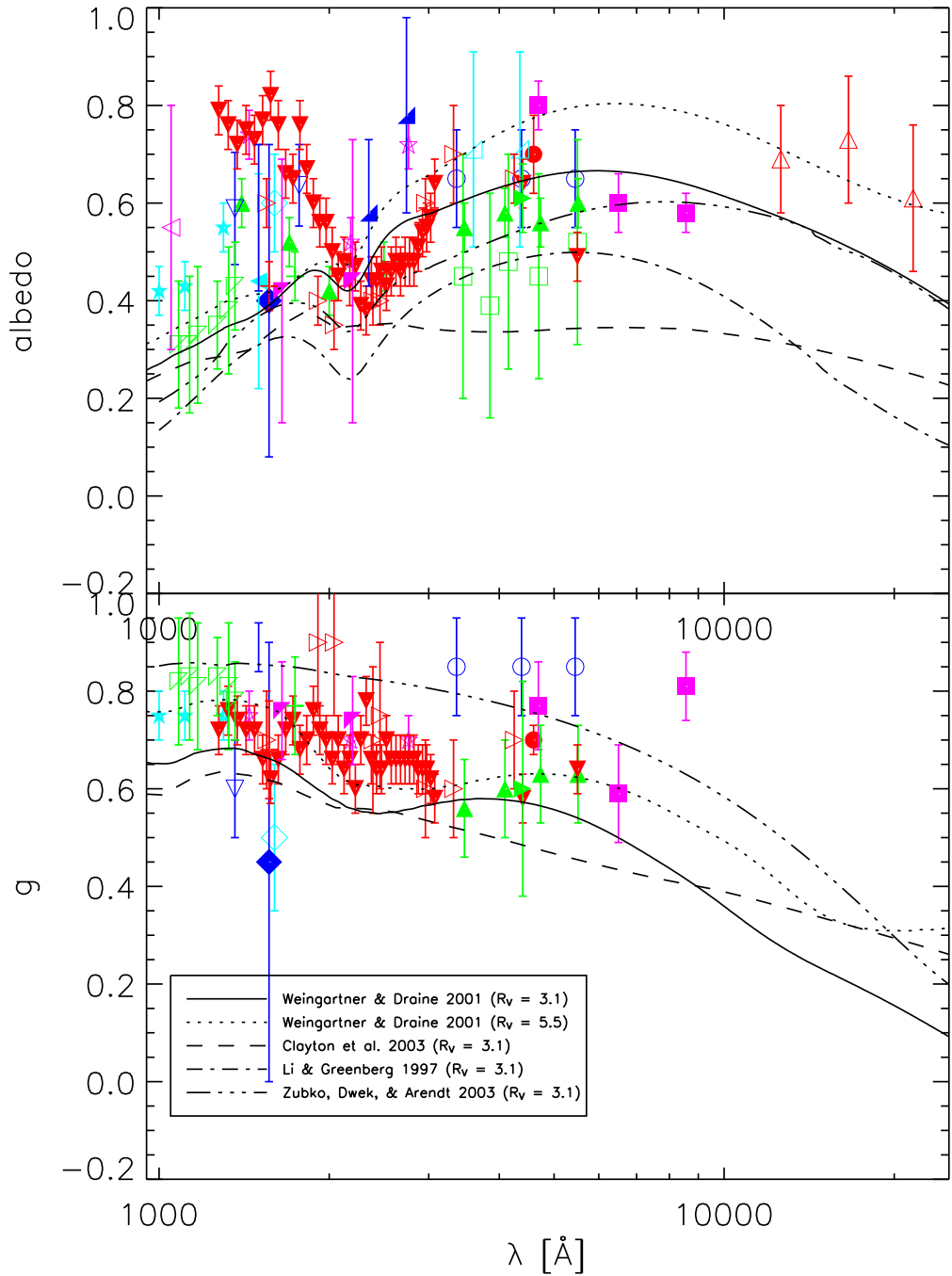}
\caption{The combined determinations of the albedo and g in reflection
nebulae, dark clouds, and the DGL are plotted versus wavelength.  The
plot symbols are not identified and do not correspond to legends in
previous figures, see Figs.~\ref{fig_rn_vals}-\ref{fig_dgl_vals} for
information about specific studies.  In addition, predictions from
dust grain models are plotted for comparison. \label{fig_all_vals}}
\end{figure}

The studies satisfying the first three criteria are listed in
Table~\ref{tab_litstudies}, separated by astrophysical object studied.
These tables give the study reference, a brief description, a model
code, and whether the study satisfied the fourth criteria and, as
such, was included in Figs.~\ref{fig_rn_vals}-\ref{fig_all_vals}.
Each model code is explained in Table~\ref{tab_models} where the
method of calculation, sources, and dust geometry are summarized.
Ideally, the model used to interpret the observations of a specific
object would include realistic illuminators (location and spectrum), a
realistic dust distribution, and calculate the full multiple
scattering of photons.  This ideal is unlikely to be fully met by any
specific model, but models which come closest have the best chance of
producing good dust scattering properties.  The original study
reference should be consulted for the full details of each study.

The $a$ and $g$ values for the studies which satisfy all four criteria
are plotted separate in Figs.~\ref{fig_rn_vals}-\ref{fig_dgl_vals} for
reflection nebulae, dark clouds, and the DGL.  In addition,
predictions for different dust grain models are included in these
figures (see \S\ref{sec_model} for model details).  The results have
been divided between object classes to highlight their differences.
It would not be surprising to find real differences in the $a$ and $g$
values between object classes.  For example, reflection nebulae and
dark clouds possess, on average, larger dust grains than the DGL.
This is seen from the different $R_V$ values measured for these
objects as $R_V$ is a rough measure of the average grain size.
Examining the results separately also allows for the impact of the
different strengths and weaknesses of each object on the derived
scattering parameters to be examined.  The final reason to separate
the results is to check if the assumptions in the modeling based on
object type significantly affect the resulting scattering values.  All
of the $a$ and $g$ values have been plotted together in
Fig.~\ref{fig_all_vals} to check the consistency of determinations
between the three astrophysical object types.

\begin{table}[tbp]
\caption{Literature Dust Scattering Studies \label{tab_litstudies}}
\begin{tabular}{llcl} \tableline\tableline
Reference & Description & Model & Inc.\ \\ \tableline
\multicolumn{4}{c}{Reflection Nebulae} \\ \tableline
Witt et al.\ 1982 & NGC 7023; IUE 1200--3000 \AA\ & RN1 & Yes$^1$ \\
 & \& ground-based 3470--5515 \AA\ & & \\
Witt et al.\ 1992 & NGC 7023; UIT 1400 \& 2800 \AA\ & RN1 & Yes \\
Witt et al.\ 1993 & NGC 7023; & RN1 & Yes \\
 & Vogager 2 1000--1300 \AA\ & & \\
Gordon et al.\ 1994 & Sco OB2; 1365 \& 1769 \AA\ & RN2 & Yes \\
Calzetti et al. 1995 & IC 435; IUE 1200--3100 \AA\ \& & RN1 & Yes \\
 &  ground-based B \& V & & \\
Burgh et al.\ 2002 & NGC 2023; FOT 900--1400 \AA\ & RN1 & Yes \\
Gibson et al. 2003 & Pleiades; WISP 1650 \& 2200 \AA\ & RN3 &
  Yes \\ \tableline
\multicolumn{4}{c}{Dark Clouds} \\ \tableline
Matilla 1970 & Coalsack and Libra dark & DC1 & Yes \\
 & cloud; UBV & & \\
Fitzgerald et al. 1976 & Thumbprint Nebula; B & DC1 & Yes \\
Laureijs et al. 1987 & L1642; 3500--5500 \AA\ & DC1 & Yes$^2$ \\
Witt et al. 1990 & Bok globule; 4690--8560 \AA\ & DC1 & Yes \\
Hurwitz 1994 & Taurus molecular cloud; & DC2 & Yes \\
 & Berkeley UVX 1600 \AA\ & & \\
Haikala et al. 1995 & G251.2+73.3 cirrus cloud; & DC1 & Yes \\
 & FAUST 1400--1800 \AA\ & & \\
Lehtinen \& Mattila 1996 & Thumbprint Nebula; JHK & DC1 & Yes$^2$ \\ \tableline
\multicolumn{4}{c}{Diffuse Galactic Light} \\ \tableline
Witt 1968 & ground-based 3600, 4350, & DGL1 & No$^3$ \\
 & \& 6100 \AA\ & & \\
Mathis 1973 & ground-based 3600 \& & DGL2 & Yes$^2$ \\
 & 4350 \AA\ & & \\
Witt \& Lillie 1973 & OAO-2 1500--4200 \AA\ & DGL3 & No$^4$ \\
Lillie \& Witt 1976 & OAO-2 1500--4200 \AA\ & DGL2 & Yes \\
Morgan et al. 1976 & TD-1 2350 \& 2740 \AA\ & DGL3 & Yes$^2$ \\
Toller 1981 & Pioneer 10 4400 \AA\ & DGL4 & Yes \\ 
Hurwitz et al. 1991 & Berkeley UVX 1625 \AA\ & DGL5 & No$^5$ \\
Murthy et al. 1993 & Voyager 2 1050 \AA\ & DGL6 & Yes$^2$ \\
Murthy \& Henry 1995 & Berkley UVX \& others & DGL6 & Yes$^2$ \\
Sasseen \& Deharveng 1996 & FAUST 1565 \AA\ & DGL7 & No$^6$ \\
Petersohn 1997 & DE 1 1565 \AA\ & DGL8 & Yes \\
Witt et al. 1997 & FAUST 1564 \AA\ & DGL8 & Yes \\
Schiminovich et al. 2001 & NUVIEWS 1740 \AA\ & DGL9 & Yes \\
\tableline\tableline
\end{tabular} \\
$^1$UV $g$ determinations not included, not enough radial data for
unique $g$ solution \\
$^2$No $g$ determination possible \\
$^3$Superseded by Mathis 1973 \\
$^4$Superseded by Lillie \& Witt 1976 \\
$^5$Superseded by Murthy \& Henry 1995 \\
$^6$Superseded by Witt et al. 1997 
\end{table}

\begin{table}[tbp]
\caption{Radiative Transfer Models \label{tab_models}}
\begin{tabular}{lccc} \tableline\tableline
Name & Method & Sources & Dust Geometry \\ \tableline
RN1 & Monte Carlo & single star & homogeneous sphere \\
RN2 & Monte Carlo & multiple stars & homogeneous sphere \\
RN3 & analytic & multiple stars & approximated \\
  & single scattering & & clumpy slab \\
DC1 & Monte Carlo & MW ISRF \& & homogeneous sphere \\
 & & specific stars & \\
DC2 & numerical int. & MW ISRF & 2D distribution \\
DGL1 & analytic & Galaxy model & homogeneous slab \\
DGL2 & numerical int. & MW ISRF & infinite cylinder \\
 & n=8 scatterings & boundary condition & \\
DGL3 & analytic & constant & infinite slab \\
DGL4 & numerical int. & star counts & non-homogeneous \\
  & n=2 scatterings & & \\
DGL5 & numerical int. & TD-1 star catalog & clumped dust \\
DGL6 & numerical int. & SKYMAP catalog & dust based on \\
 & & & HI survey \\
DGL7 & Monte Carlo & star catalogs & dust based on \\
 & & & HI survey \\
DGL8 & Monte Carlo & TD-1 star catalog & dust cloud spectrum \\
 & & & based on HI survey \\
DGL9 & Monte Carlo & 3D TD-1 star catalog & dust cloud spectrum \\
 & & & based on HI survey \\
\tableline\tableline
\end{tabular}
\end{table}

\section{Discussion}

The brightness and relative compactness of reflection nebulae have
led to determinations of dust scattering properties with a finer
wavelength sampling and smaller uncertainties than in either dark
clouds or the DGL (see Fig.~\ref{fig_rn_vals}).  Due to the usual
presence of strong emissions (ERE and non-equilibrium emission),
reflection nebulae have only provided scattering information in the
ultraviolet and blue-optical.  While dark clouds are faint, their
relative compactness and usual lack of strong emissions has made them
valuable as probes of dust scattering properties in the red-optical
and near-infrared (Fig.~\ref{fig_dc_vals}).  The one drawback to both
reflection nebulae and dark cloud studies is that they probe higher
density media than diffuse interstellar medium.  This leads to
questions to the general applicability of reflection nebulae and dark
clouds results.

The faintness and large extent of the DGL makes it the most
challenging object to both observe and model.  The importance of
measuring dust scattering properties in the diffuse interstellar
medium and determinations of the extragalactic background has resulted
in considerable effort being expended on the study of the DGL.
Determinations of DGL dust scattering properties in the blue-optical
and near-ultraviolet have been fairly uncontroversial, while such
determinations in the far-ultraviolet have been quite controversial.
This controversy is highlighted by the back-to-back reviews on the
ultraviolet background in which it was argued that the majority of the
ultraviolet background is from dust scattered light (Bowyer 1991) or
extragalactic light (Henry 1991).  The difficulties in both observing
and modeling the DGL in the far-ultraviolet has led to studies in the
literature with conflicting results, either low $a$ and $g$ values or
high $a$ and $g$ values.  This is the area which was most affected by
the fourth criteria (not superseded by a more recent study).  A number
of initial analyses giving low $a$ and $g$ values were superseded by
subsequent studies using more sophisticated models which found higher
$a$ and $g$ values.  As a result, the scatter in the plots of $a$ and
$g$ for the DGL (Fig.~\ref{fig_dgl_vals}) is now within the quoted
uncertainties.  The values of $a$ and $g$ derived from DGL
measurements are similar to those derived from reflection nebulae and
dark clouds.

By examining Figs.~\ref{fig_rn_vals}-\ref{fig_dgl_vals}, it can be
seen that the $a$ and $g$ values derived for each object type are
internally consistent.  For example, the measurements of $a$ and $g$
from different reflection nebulae all are consistent with a single
wavelength dependence.  When examing these figures, it is important to
remember that they include results from broad-band filters along with
those derived from spectroscopy.  Comparing the results for all three
object types in Fig.~\ref{fig_all_vals}, it can be seen that the same
may be true in general and not just internal to an object type.  The
only area where the data from different object types may disagree is
in the far-ultraviolet region for $a$ measurements only.  The
uncertainties are large enough to make a strong statement either way
difficult.  The high level of agreement between object types implies
that their different strengths and weaknesses as well as model
assumptions are not severely biasing our view of dust scattering
properties.  This result goes a long way in answering the concerns
raised by Mathis, Whitney, \& Wood (2002) about possible biases in
reflection nebulae measurements of the dust albedo caused by ignoring
the known clumpiness of dust.

In fact, it is not unreasonable to say that the current data (see
Fig.~\ref{fig_all_vals}) indicates the wavelength dependence of the
albedo is $\sim$0.6 in the near-infrared/optical with a dip to
$\sim$0.4 for 2175~\AA\ bump, a rise to $\sim$0.8 around 1500~\AA, and
a drop to $\sim$0.3 by 1000~\AA.  The wavelength dependence of $g$ is
simpler with a monotonic rise from 0.6 to 0.8 from 10000 to 1000~\AA.
The uncertainties do allow for significant real variation in the $a$
and $g$ values around these qualitative averages of at least 0.1.

The study of dust scattering properties has shed light on the nature
of the 2175~\AA\ bump and the far-ultraviolet rise features of
extinction curves.  Early work on the 2175~\AA\ bump indicated that it
was likely an absorption feature with no scattered component (eg.,
Lillie \& Witt 1976).  Subsequently, evidence for a scattered
component in the 2175~\AA\ bump was found in two reflection nebulae,
CED 201 and IC 435, by Witt, Bohlin, \& Stecher (1986).  This results
was found to definitely be spurious for IC 435 when much more
sensitive observations were taken and analyzed by Calzetti et al.\
(1995).  By inference, the observations of CED 201 are also likely to
be spurious.  As a result, all evidence currently supports a 2175~\AA\
extinction bump which is only due to absorption.  Similarly, examining
the results of reflection nebulae (see Fig.~\ref{fig_rn_vals}) gives
good evidence that a significant portion of the far-ultraviolet rise
in extinction curves (1700~\AA\ and shorter wavelengths) is due to
absorption.

\subsection{Comparisons with Dust Grain Models \label{sec_model}}

One way to synthesize our knowledge of dust is to fit different
observations of dust to a dust grain model.  The main observational
constraints for dust grain models are usually the diffuse ISM dust
extinction curve and dust abundances (Weingartner \& Draine 2001;
Clayton et al.\ 2003).  In addition, some dust grain models also fit
the dust polarization (Kim \& Martin 1994; Li \& Greenberg 1997) and
infrared emission spectrum (Zubko et al. 2003; Li \& Draine 2001).
Some dust grain models also have been fit to the denser dust
extinction curves (Kim, Martin, \& Hendry 1994; Weingartner \& Draine
2001).  No dust grain models have used the measured dust scattering
properties as fitting constraints, but most do use the scattering
properties as a consistency check.

In Figs.~\ref{fig_rn_vals}-\ref{fig_all_vals}, predictions for $a$ and
$g$ from recent dust grain models are plotted.  The Zubko et al.\
(2003) model plotted is described as BARE-NC-B.  All but one of the
five model predictions plotted are for diffuse ISM dust ($R_V = 3.1$)
and are then only directly comparable to the DGL dust scattering
results.  As can be seen from Fig.~\ref{fig_dgl_vals}, the dust grain
models are in fairly good agreement with the results for $g$, but most
of the models underpredict $a$ for most wavelengths.  All the models
have a small dip in $a$ at the 2175~\AA\ bump, but the dip is not as
large as indicated by the DGL measurements.  The models all predict
lower far-ultraviolet albedos than have been measured in the DGL.

The one dust grain model for denser dust (Weingartner \& Draine; $R_V
= 5.5$) is directly comparable to the dust scattering measurements in
reflection nebulae (Fig.~\ref{fig_rn_vals}) and dark clouds
(Fig.~\ref{fig_dc_vals}) as these objects have similar $R_V$ values.
Similar to the DGL comparison, the $g$ values predicted by the model
are in reasonable agreement for reflection nebulae and dark clouds,
except possible in the optical for dark clouds.  The model predictions
for $a$ agree with the measurements for dark clouds which mainly probe
the optical and near-infrared.  The $a$ predictions for reflection
nebulae are generally too high in the optical and too low in the
far-ultraviolet.  The albedo dip at 2175~\AA\ is much smaller in the
model than the well measured values indicate.  Basically, the recovery
of the albedo at around 1500~\AA\ is not seen in any of the dust grain
models.

\section{Summary and Future}

This review of dust scattering determinations has concentrated on
giving a critical examination of results presented in the literature
for reflection nebulae, dark clouds, and the DGL.  Unlike previous
such reviews, each study was required to pass four simple criteria for
inclusion in this review.  There were a total of 23 studies passing
the four criteria with 7 on reflection nebulae, 7 on dark clouds, and
9 on the DGL.

A great deal of progress has been made since the last review of this
area (Witt 1989).  The uncertainty in far-ultraviolet albedo from DGL
studies has been resolved with the help of better observations and
more sophisticated modeling.  The 2175~\AA\ bump has been shown to be
an absorption feature; earlier indications of scattering in the bump
have been traced to bad data.  The measurements in reflection nebulae,
dark clouds, and the DGL are roughly consistent alleviating earlier
worries about the effects of the modeling assumptions specific to the
object being studied (eg. reflection versus the DGL).

While much progress has been made in measuring dust scattering
properties, there are a number of questions which are outstanding.

\begin{enumerate}

\item Does the assumption of the Henyey-Greenstein single parameter
scattering phase function significantly bias the resulting dust
scattering parameters ($a$ and $g$)?  This question is probably best
answered with models of images of reflection nebulae and dark clouds
which include either more complicated analytical phase functions (eg.,
a double Henyey-Greenstein phase function; eq.~4 of Witt 1977) or
scattering phase functions computed from dust grain models.

\item What are the $a$ and $g$ values in the near-infrared?
Currently, there is only $a$ measurements for a single dark cloud and
no $g$ measurements.  The $a$ and $g$ values in the near-infrared
probe the larger dust grains, information about which is difficult to
determine from extinction curve measurements alone.

\item What are the $a$ and $g$ values in the red-optical and
near-infrared for reflection nebulae?  Is it even possible to
determine $a$ and $g$ in reflection nebulae for these wavelengths?
There are reflection nebulae in which ERE has not been detected (Witt
\& Boroson 1990).  Are there nebulae without near-infrared,
non-equilibrium emission?

\item Is it possible to measure dust scattering properties in the
red-optical and near-infrared from DGL measurements?

\item What does it mean to measure dust scattering properties in
objects without ERE or near-infrared, non-equilibrium emission if the
diffuse ISM has been shown to have ERE (Gordon, Witt, \& Friedmann
1998) and probably also has near-infrared, non-equilibrium emission.

\item Is it possible to use the full multiwavelength appearance of
reflection nebulae and dark clouds to alleviate lingering concerns
about the geometrical assumptions inherit in modeling such objects?
Specifically, the inclusion of ultraviolet through far-infrared would
provide direct measurements of the direct, scattered, and re-emitted
light and thus require many fewer assumptions.

\item Is it feasible to consider using the full multiwavelength
appearance of the DGL to reduce number of assumption necessary to
model the DGL?  Or is this still too complex of a problem, especially
in light of the need for accurate 3D star positions of all stars
important for the DGL (eg., hot stars in the ultraviolet and cooler
stars at longer wavelengths)?

\item What are the real differences between dust scattering
properties in reflection nebulae, dark clouds, and the DGL?  The
current evidence indicates they are consistent with each other at the
$\sim$20\% level.

\item Are the dust scattering properties in other galaxies different
than those derived for Milky Way dust?  The Large and Small Magellanic
Clouds offer environments which have lower metallicities and high star
formation rates than the Milky Way.  In addition, dust extinction
curves in the Magellanic Clouds have been measured to be quite
different from Milky Way extinction curves (Gordon et al. 2003).

\end{enumerate}

Finally, the author would like to encourage readers who find studies
which have not been included in this review to email them to the
author (currently kgordon@as.arizona.edu).  The author is committed to
continuing to update the web-based version of this review\footnote{currently
at http://dirty.as.arizona.edu/$\sim$kgordon/Dust/Scat\_Param/scat\_data.html}.

\end{document}